\documentclass[sn-mathphys]{sn-jnl}

\usepackage{graphicx}
\usepackage{bm}
\usepackage{amsbsy}
\usepackage{amsmath}
\usepackage{amssymb}
\usepackage{xcolor}   
\usepackage{latexsym}

\jyear{2023}%

\newcommand{\dd}{\mathrm{d}}			
\newcommand{\bF}{\mathbf{f}}			
\newcommand{\bg}{\mathbf{g}}			
\newcommand{\bh}{\mathbf{h}}			
\newcommand{\bH}{\mathbf{H}}			
\newcommand{\bK}{\mathbf{K}}			
\newcommand{\bM}{\mathbf{M}}			
\newcommand{\bp}{\mathbf{p}}			
\renewcommand{\br}{\mathbf{r}}			
\newcommand{\bR}{\mathbf{R}}			
\newcommand{\bu}{\mathbf{u}}			
\newcommand{\bv}{\mathbf{v}}			

\newcommand{\bV}{\mathbf{V}}
\newcommand{\bW}{\mathbf{W}}			
\newcommand{\tauB}{\tau_{\rm B}}		
		
\newcommand{\bsigma}{\boldsymbol{\sigma}}
\newcommand{\bomega}{\boldsymbol{\omega}}
\newcommand{\bOmega}{\boldsymbol{\Omega}}
\newcommand{\bnabla}{\boldsymbol{\nabla}}
\newcommand{\kT}{k_{\rm B}T}			
\newcommand{\sK}{s_{\rm K}}			%
\newcommand{\ellp}{\ell_{\rm p}}
\newcommand{\mylambda}{r}			

\newcommand{\fext}{f^{\rm ext}}		

\newcommand{\ave}[1]{\langle #1 \rangle}

\usepackage{booktabs}

\begin{document}

\title[Magneto-Permeability in Ferrofluid Flow through Porous Media]{Magneto-Permeability Effect in Ferrofluid Flow through Porous Media studied via Multiparticle Collision Dynamics}

\author*[1]{\fnm{Patrick} \sur{Ilg}}\email{p.ilg@reading.ac.uk}

\affil*[1]{\orgdiv{School of Mathematical, Physical, and Computational Sciences}, \orgname{University of Reading}, \orgaddress{\city{Reading}, \postcode{RG6 6AX}, \country{United Kingdom}}}

\abstract{As more and more promising applications of magnetic nanoparticles in complicated environments are explored, their flow properties in porous media are of increasing interest. 
We here propose a hybrid approach based on the Multiparticle Collision Dynamics Method extended to porous media via friction forces and coupled with Brownian Dynamics simulations of the rotational motion of magnetic nanoparticles' magnetic moment. 
We simulate flow in planar channels homogeneously filled with a porous medium and verify our implementation by reproducing the analytical velocity profile of the Darcy-Brinkman model in the non-magnetic case. 
In the presence of an externally applied magnetic field, 
the non-equilibrium magnetization and friction forces lead to field-dependent velocity profiles that result in effective, field-dependent permeabilities. 
We provide a theoretical expression for this magneto-permeability effect in analogy with the magneto-viscous effect. 
Finally, we study the flow through planar channels, where only the walls are covered with a porous medium. We find a smooth crossover from the Poiseuille profile in the center of the channel to the Brinkman-Darcy flow in the porous layers. We propose a simple estimate of the thickness of the porous layer based on the flow rate and maximum flow velocity.}


\maketitle

\section{Introduction}

Dynamics and flow of magnetic nanoparticles (MNPs) suspended in non-magnetic viscous carrier fluids (ferrofluids) have been studied intensively over the last decades \cite{socoliuc_ferrofluids_2022,felicia_recent_2016,Ilg_lnp,rosensweigbook}. 
These studies have almost exclusively focused on spatially homogeneous solvents. 
In many applications, however, one is interested in the flow through composite materials. 
Prominent examples are the transport through sand or other granular matter \cite{Sahimi_porous}. 
On a coarse-grained level, transport phenomena through such materials can be described via a porous medium approach  \cite{Sahimi_porous,plessis_flow_1991}. 
Similarly, the porous medium approach is also frequently used in biomedical applications to describe transport through biological tissues and cancerous cells (see e.g.\ \cite{khaled_role_2003,tucci_pennes_2021,al_sariri_multi-scale_2022,mosharaf-dehkordi_fully_2019} and references therein).

Although MNPs find more and more promising biomedical and technical applications \cite{socoliuc_ferrofluids_2022,Parak_review,felicia_recent_2016}, to date, only a handful of studies address the flow properties of MNPs through porous media. 
One of the experimental studies on ferrofluid flow though sands and sediments observed a strong dependence on external magnetic fields  \cite{borglin_experimental_2000}. 
Similarly, experiments on the internal convection of ferrofluids flowing through a capillary tube filled with porous media showed that external fields could significantly enhance the thermal conductivity \cite{shafii_experimental_2018}. 
Also the efficiency of ferrofluids for oil displacements in a sand-filled pipe was investigated experimentally and compared to finite-element simulations \cite{dou_oil_2022}. 

Fluid dynamics simulation of ferrofluids have also been performed to investigate their use as displacing fluid in fractured porous media \cite{huang_numerical_2017}. 
Other finite-element or finite-volume simulations have addressed some particular flow \cite{huang_numerical_2021,guerroudj_ferrohydrodynamics_2023} and heat transfer \cite{abbas_flow_2021} properties of ferrofluid flow through porous media. 
A porous medium approach was also used in finite-volume simulations of magnetic drug targeting of MNPs, coupling channel flow to adjacent tumor region via the permeable endothelium layer   
\cite{nemati_numerical_2017}. 
These simulation studies relied on highly simplified constitutive models, typically neglecting internal rotations and corresponding non-equilibrium magnetization components.

In addition to classical fluid dynamics simulations such as finite-volume and finite-element methods, the Lattice Boltzmann scheme has been successfully used to describe flow through porous media  \cite{dardis_lattice_1998}. 
In these simulations, explicit scatterers for fluid motion are placed at fixed locations within the simulation cell. 
Using this method, ferrofluid permeation into a randomly structured porous medium has been simulated and shown to be sensitive to an applied magnetic field \cite{hadavand_ferrofluid_2013}. 
As an alternative simulation approach, an extension of the highly versatile multi-particle collision dynamics method (MPC) \cite{malevanets_mesoscopic_1999} to transport in porous media has been proposed in Ref.\ \cite{matyka_sedimentwater_2017} for non-magnetic fluids. 
In the latter, the effect of porous media on fluid transport is simply modelled as local damping, leading to very efficient simulation methods. 
Note that the Lattice Boltzmann methods put forward in Refs.\ \cite{dardis_lattice_1998,hadavand_ferrofluid_2013} resolve the detailed fluid dynamics in the vicinity of individual grain boundaries. On the other hand, the MPC modelling proposed in Ref.\ \cite{matyka_sedimentwater_2017} is suitable for a more coarse-grained level of description where the porous medium can be considered to be locally homogeneous. 

Here, we use similar ideas to simulate ferrofluid flow through porous media via a MPC method that is coupled to Brownian Dynamics simulations of the MNP dynamics to model their internal rotations.  
In the absence of porous media, this approach was proposed and validated in Ref.\ \cite{ilg_multiparticle_2022}, showing the correct incorporation of a reliable constitutive model for dilute ferrofluids. 
We here show how this model can be extended via friction forces to model ferrofluid flow through porous media. 
In the absence of an external magnetic field, we reproduce the Darcy-Brinkman velocity profile and clarify the interpretation of and relationship between the model parameters. 
Simulations of driven flow through planar channels show that flow properties can be manipulated by external magnetic fields. 
In particular, we observe an effective permeability that increases with increasing field strength before reaching a limiting value. 
Using kinetic theories of ferrofluids, we provide a theoretical expression of this magneto-permeability effect in close analogy with the magneto-viscous effect. 
Finally, we study driven flow through planar channels where only the channel walls are covered with a layer of porous material, with no porous medium present in the center region of the channel.

\section{Modeling}

\subsection{Continuum Level} \label{continuum.sec}

Darcy's law predicts the flow velocity $\bv$ through a porous medium when a pressure gradient $\bnabla p$ is applied as \cite{Sahimi_porous}
\begin{equation} \label{Darcy}
\bv = - \frac{1}{\eta}\bK\cdot\bnabla p,
\end{equation}
where $\eta$ is the dynamic viscosity of the fluid. 
The empirical proportionality coefficient $\bK$ is known as permeability of the porous medium.  
For isotropic porous media $\bK=K{\bf I}$, with ${\bf I}$ the identity matrix, so that $\bv = -(K/\eta)\bnabla p$. 
In a finite domain, the
Darcy-Brinkmann model provides a better description than Darcy's law \cite{khaled_role_2003}. 
This model can be formulated as the stationary Navier-Stokes equation supplemented with an additional damping term  proportional to an empirical parameter $\alpha$, 
\begin{equation} \label{DarcyBrinkman}
\nu \nabla^2 \bv - \alpha \bv = \frac{1}{\rho}\bnabla p. 
\end{equation}
The density and kinematic viscosity of the fluid are denoted by $\rho$ and $\nu=\eta/\rho$, respectively. 
 The phenomenological parameter $\alpha$ governs the strength of the damping term and is related to the permeability coefficient by $\alpha=\nu/K$. 
We here consider Reynolds numbers that are small enough so that the Forchheimer correction  \cite{khaled_role_2003}  is irrelevant. 

For spatially homogeneous porosity, i.e.\ a position-independent $\alpha$, the exact solution of Eq.\ \eqref{DarcyBrinkman} for one-dimensional channel flow $\bv(\br)=v(y){\bf e}_x$ reads  \cite{dardis_lattice_1998}
\begin{equation} \label{v_channel}
v(y) = c \left( 1 - \frac{\cosh(\mylambda [y-L/2])}{\cosh(\mylambda L/2)} \right),
\end{equation}
where $L$ denotes the width of the channel and no-slip boundary conditions on the channel walls have been assumed. 
The parameters appearing in the flow profile Eq.\ \eqref{v_channel} are given by 
 $c=-(\alpha\rho)^{-1}\frac{\dd p}{\dd x}$ and 
\begin{equation} \label{lambdasq}
\mylambda^2 = \alpha / \nu.
\end{equation}
From Eq.\ \eqref{lambdasq} and the above relation $\alpha=\nu/K$, we find that the permeability can also be expressed as 
$K=1/\mylambda^2$. These relations will be useful for later analysis. 

Note that for weak damping, we recover the usual Poiseuille profile from Eq \eqref{v_channel}, $v(y)=-(2\nu\rho)^{-1}\frac{\dd p}{\dd x}y(L-y) + {\cal O}(\mylambda^2)$. 
Conversely, increasing $\mylambda$ leads to stronger and stronger deviations from the parabolic velocity profile. 

The amount of fluid transported per unit time through a cross-section of the channel (known as volumetric flow rate in the three-dimensional case) is obtained from $\dot{\mathcal{V}}=\int_0^Lv_x(y)\dd y$. 
For the velocity profile \eqref{v_channel} we obtain 
$\dot{\mathcal{V}}=cL[1-\tanh(L^\ast)/L^\ast]$ with reduced channel width $L^\ast=\mylambda L/2$. 
As expected, the flow rate is proportional to the applied pressure gradient. 
For $L^\ast\gg 1$ we find $\dot{\mathcal{V}}\approx c[L-2/\mylambda]$ corresponding to a plug flow, whereas for 
$L^\ast\ll 1$ we recover the equivalent of the Hagen-Poiseuille law  for two-dimensional channels, 
$\dot{\mathcal{V}}\approx - (\dd p/\dd x) L^3/[12\rho\nu] ( 1 + {\cal O}((L^\ast)^2) )$. 

\subsection{Mesoscopic Level: Particle-based model} \label{MPCbasics.sec}

Over the past decades, several particle-based methods for simulating fluid flow have been explored (see e.g.\ \cite{noguchi_particle-based_2007} and references therein). 
Contrary to more traditional fluid dynamics simulations, these mesoscopic methods are very flexible, straightforward to implement, and naturally include thermal fluctuations. 
In the present study, we employ one of these methods known as multi-particle collision dynamics (MPC) \cite{malevanets_mesoscopic_1999}. 
One of the advantages of the MPC method is that viscoelastic fluids can be modeled rather straightforwardly \cite{Gompper:2009is}. 
In particular, we have already proposed and successfully tested an MPC implementation of ferrofluid flow using a reliable constitutive model \cite{ilg_multiparticle_2022}. 
To make the paper self-contained, we provide a short description of the standard MPC method, before specifying the extension to porous media. 

Within the MPC method, the fluid is represented by a collection of $N$ identical particles, each with mass $m$. 
If $\br_i$ and $\bv_i$ denote the position and velocity of particle $i$, $i=1,\ldots,N$, particle dynamics is split into a streaming and a collision step. 
In the streaming step, particles are advanced for a time $\Delta t$ as
\begin{align}
\br_i(t+\Delta t) & = \br_i(t) + \bv_i(t)\Delta t + \frac{\Delta t^2}{2m}\bF_i(t) \label{r_streaming}\\
\bv_i'(t) & = \bv_i(t) + \frac{\Delta t}{m}\bF_i(t) , \label{v_streaming}
\end{align} 
where $\bF_i(t)$ is the total force acting on particle $i$ at time $t$. 
While Eqs.\ \eqref{r_streaming},\eqref{v_streaming} are formally identical to those used in Molecular Dynamics simulations, 
the crucial idea of MPC as a mesoscopic method is to include in $\bF_i$ only external and body forces and to disregard inter-particle interactions in the streaming step. 
Instead, inter-particle interactions are accounted for via momentum exchange in the collision step. 
In this collision step, all particles $i$ residing at time $t$ in the same collision cell $C_i$ are updated simultaneously as    
\begin{equation} \label{collision}
\bv_i(t+\Delta t) = \bV_{C_i}(t) + \beta_{\rm th}\bR\cdot[\bv_i'(t) - \bV_{C_i}(t)]. 
\end{equation}
In Eq \eqref{collision}, $\bV_{C_i}(t)$ denotes the center-of-mass velocity of the collision cell $C_i$ and $\bR=\bR(\alpha)$ a matrix, describing rotations around a randomly chosen axis by an angle $\pm\alpha$. 
Equation \eqref{collision} models the effect of collisions among particles as rotations of their relative velocities. 
A local thermostat is present in Eq.\ \eqref{collision} and described by the factor $\beta_{\rm th}=\sqrt{T/T_{C_i}}$, where 
$T_{C_i}$ is the instantaneous kinetic temperature of the collision cell $C_i$ and $T$ a prescribed bath temperature. 
We use a two-dimensional square grid, so the collision cells are squares of side length $a$. 
Using a spatially fixed grid of collision cells violates Galilean invariance. Therefore, we follow common practice \cite{Ihle:2001bf} and in each step shift the grid of collision cells by a vector with independent random components in $[-a/2,a/2]$.

Besides the time step $\Delta t$ which determines the mean-free path $\lambda=\Delta t \sqrt{\kT/m}$, the mean number of particles per collision cell $Q$ is the other crucial  parameter in the MPC model \cite{malevanets_mesoscopic_1999,noguchi_transport_2008}.
Due to the importance of angular momentum conservation, we here follow our earlier work \cite{ilg_multiparticle_2022} and implement the angular momentum-conserving algorithm (denoted as MPC-DR in Ref.\ \cite{noguchi_transport_2008}), where the rotation angle $\alpha$ is not a free parameter but chosen as 
\begin{equation}
\cos\alpha = \frac{1-R^2}{1+R^2},
\end{equation} 
where $R=A_1/A_2$ with angular 
$A_1=\sum_{j\in C_j}[\br_j\times\tilde{\bv}_j]_z$ and projected
$A_2=\sum_{j\in C_j}\br_j\cdot\tilde{\bv}_j$ relative velocities before collision,  
$\tilde{\bv}_j=\bv'_j-\bV_{C_j}$. 

The simplified collision rules between particles \eqref{collision} do not resolve individual collisions, 
but ensure local conservation laws are obeyed. 
Therefore, the MPC method is numerically very efficient and hence can simulate hydrodynamic behavior on time and length scales larger than $\Delta t$ and $a$  \cite{malevanets_mesoscopic_1999,noguchi_particle-based_2007}. 

In the form described so far, the MPC method has been successfully applied to various flow problems for viscous and viscoelastic fluids \cite{Gompper:2009is}, including 
ferrofluids \cite{ilg_multiparticle_2022}. 
In order to describe flow through porous media, however, we need to introduce the effect the porous medium exerts on the fluid via inelastic collisions. 
Here, we follow the ideas put forward in Ref.\  \cite{dardis_lattice_1998} that on a mesoscopic level, the interaction of the fluid with the porous medium can be described as a local damping of the fluid velocity. 
Within the MPC method, this approach can be implemented straightforwardly as an additional friction force 
acting on the particles 
 \cite{matyka_sedimentwater_2017}, 
\begin{equation} \label{friction}
\bF^{\rm fric}_i(t) = - \xi(\br_i(t))\bv_i(t),
\end{equation}
where $\xi(\br)$ is a (possibly position-dependent) friction coefficient. 
For $\xi=0$ we recover the original MPC model, whereas $\xi>0$ describes velocity damping due to porous media. 
For the case of pressure-drive flow that we consider in the following, the force on particle $i$ can be written as 
$\bF_i=\bF_i^{\rm fric} + \bF^{\rm ext}$, where the external force due to an applied pressure gradient is 
$\bF^{\rm ext}=-\rho^{-1}\bnabla p$. 

Eqs \eqref{r_streaming} -- \eqref{friction} describe the MPC model of non-magnetic fluid flow through porous media. 
This model has essentially been proposed in Ref  \cite{matyka_sedimentwater_2017}, where instead of adding the friction force \eqref{friction}, particle velocities are rescaled by a factor $(1-\xi\Delta t/m)$.  

\subsection{Hybrid MPC--BD model for FF flow} \label{MPCBD}

For magnetic fluids, the stationary momentum balance equation \eqref{DarcyBrinkman} must be supplemented by additional Kelvin-Helmholtz forces \cite{rosensweigbook}, 
 \begin{equation} 
\rho \bF_{\bM} = (\bM\cdot\bnabla)\bH + \frac{1}{2}\bnabla\times(\bM\times\bH), 
\label{Fmag}
\end{equation}
where $\bH$ and $\bM$ denote the magnetic field and the magnetization, respectively. 
We assume the fluid to be non-conducting, therefore we must also satisfy the magnetostatic Maxwell equations 
\begin{equation}
\bnabla\times \bH = {\bf 0}, \quad
\bnabla\cdot{\bf B} = 0, 
 \label{magnetostatics.eq}
\end{equation}
where ${\bf B}=\mu_{0}(\bH+\bM)$ denotes the magnetic induction and $\mu_0$ the permeability of free space. 
For a thorough description of ferrofluid hydrodynamics see e.g.\ Ref.\ \cite{rosensweigbook}. 

To evaluate the force density \eqref{Fmag}, one needs to employ a model to calculate the field- and flow-dependent magnetization $\bM$. 
Unfortunately, even after 50 years of research, there is no commonly agreed magnetization equation available in the literature (see e.g.\ \cite{Luecke_ZChem,fang_response_2022} and references therein). 
Therefore, we here consider dilute conditions  where there is less controversy and adopt the classical kinetic model of Martsenyuk {\em et al.}\ \cite{MRS74} that has been studied frequently since \cite{IKH01,soto-aquino_oscillatory_2011,Ilg_lnp}.  

To make the paper self-contained, we briefly present the MPC implementation of this model proposed in Ref.\ \cite{ilg_multiparticle_2022}. Further details can be found in the original reference. 
In this model, one considers the rotational dynamics of an individual MNP within the rigid-dipole approximation under the influence of external magnetic fields and flow. 
From the balance of magnetic, flow and random torques, one finds that the orientation of the magnetic moment of particle $i$ evolves to first order in the time step $\Delta t_{\rm B}$ by 
\begin{equation} 
\bu_i(t+\Delta t_{\rm B}) = 
\frac{\bu_i(t)+\Delta\bomega_i\times\bu_i(t)}{\|\bu_i(t)+\Delta\bomega_i\times\bu_i(t)\|} 
\label{Euler_u}
\end{equation}
with 
\begin{equation}
\Delta\bomega_i = \left[ \tauB\bOmega_{C_i} + \frac{1}{2} \bu_i\times\bh_{C_i} \right] \frac{\Delta t_{\rm B}}{\tauB} + \frac{1}{\sqrt{\tauB}}\Delta\bW_i, 
 \label{Euler_omega}
\end{equation}
where $\bOmega_{C_i}$ and $\bh_{C_i}$ are one half the local vorticity of the flow and the magnetic field, respectively, both evaluated at the center of collision cell $C_i$. The Brownian relaxation time of a MNP is denoted by $\tauB$ and $\Delta\bW_i$ are independent, three-dimensional Wiener increments over a time interval $\Delta t_{\rm B}$. 
In the simulations presented here, we use a weak second order stochastic Heun scheme, where Eqs.\ \eqref{Euler_u} and \eqref{Euler_omega} serve as predictor step. 

With the magnitude of the magnetic moment $\mu$ of a single MNP, the instantaneous local magnetization in collision cell $C_i$ can be calculated as 
\begin{equation}
\bM_{C_i}(t)=\frac{n\mu}{N_{C_i}(t)}\sum_{j\in C_i}\bu_j(t),
 \label{M_cell}
\end{equation} 
where 
$N_{C_i}(t)$ is the number of MPC particles in collision cell $C_i$ at time $t$ 
and $n$ denotes the number density of MNPs. 
We calculate the magnetization $\bM$ in all collision cells according to Eq.\ \eqref{M_cell} and use kernel-smoothing methods to find a discretization of the magnetization field $\bM(\br;t)$. 
From the discretized magnetization field, we calculate spatial gradients via finite-difference approximations and are thus able to evaluate the Kelvin-Helmoltz force density \eqref{Fmag} within each collision cell. 
Further details on the method are given in Ref.\ \cite{ilg_multiparticle_2022}. 

Note that we do not explicitly include the effect of the porous medium on the rotational motion of the MNPs. 
While it is plausible that inelastic collisions of fluid and MNPs with obstacles lead to an effective damping of the translational motion, their effect on rotations is less obvious, especially since we later consider rotational motion of MNPs on time scales long compare to fluid motion (by choosing $\tauB\gg\Delta t$). 
One possibility would be to model this effect as an additional rotational friction. 
In this case, all results presented below still hold when adjusting $\tauB$ correspondingly for given $\xi$. 
Due to the uncertainties associated with such modelling, we chose to consider $\tauB$ constant in this initial study.

\section{Non-magnetic fluid flow through porous media}

All numerical results and parameter values presented below are reported in dimensionless form, with the particle mass $m$ and the linear size of the collision cell $a$ as basic units, together with the reference thermal energy $\kT_{\rm ref}$. 
Consequently, the units for time are $t_{\rm ref}=a\sqrt{m/\kT_{\rm ref}}$ 
and the units for the diffusion coefficient and kinematic viscosity are $D_{\rm ref}=a\sqrt{\kT_{\rm ref}/m}$.
 
First, we consider non-magnetic fluids, i.e.\ the MPC particles experience no other external forces except friction forces \eqref{friction} and external pressure gradients, $\bF_i = \bF_i^{\rm fric} + \bF^{\rm ext}$.

\subsection{Self-diffusion in unbounded domain, no external forcing}

\begin{figure}[htbp]
\begin{center}
\includegraphics[width=0.45\textwidth]{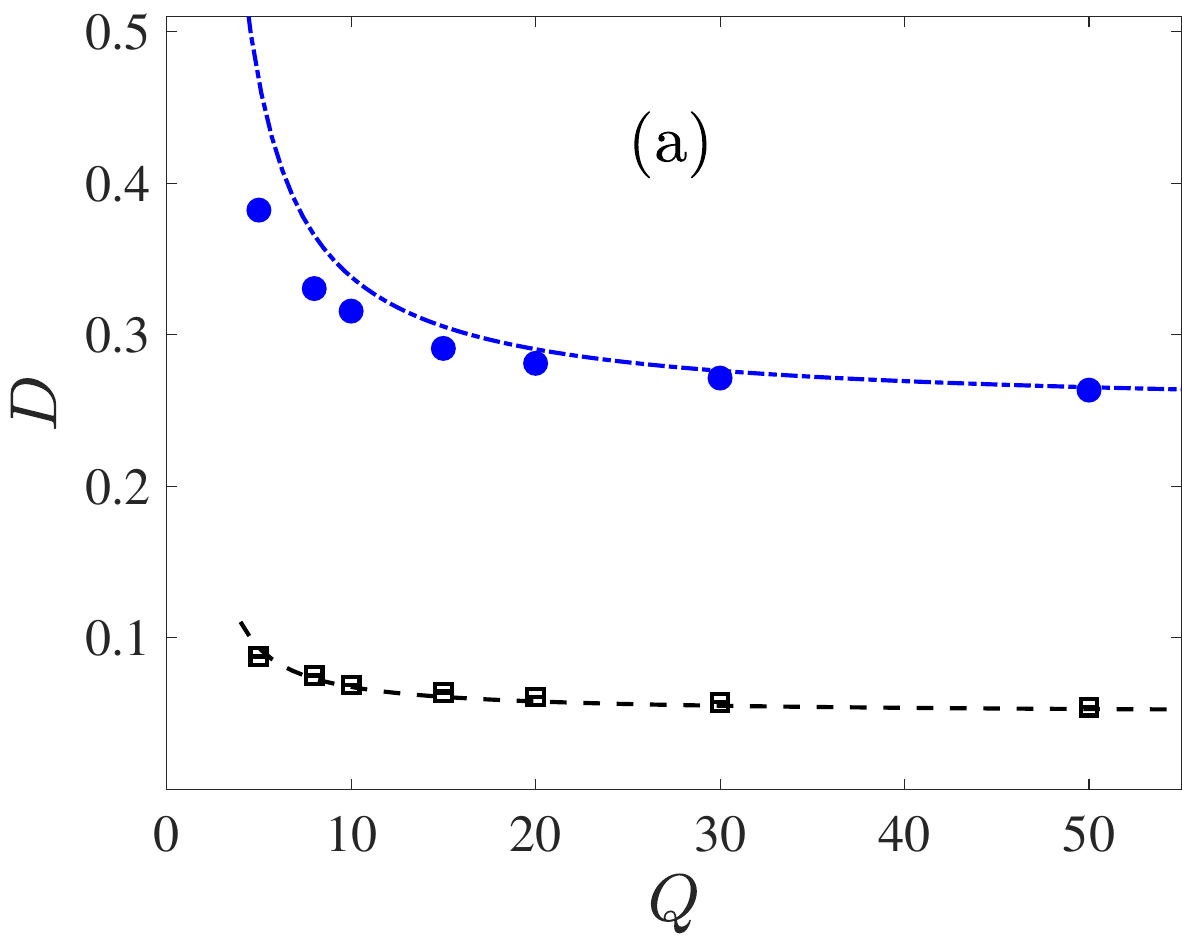}
\includegraphics[width=0.45\textwidth]{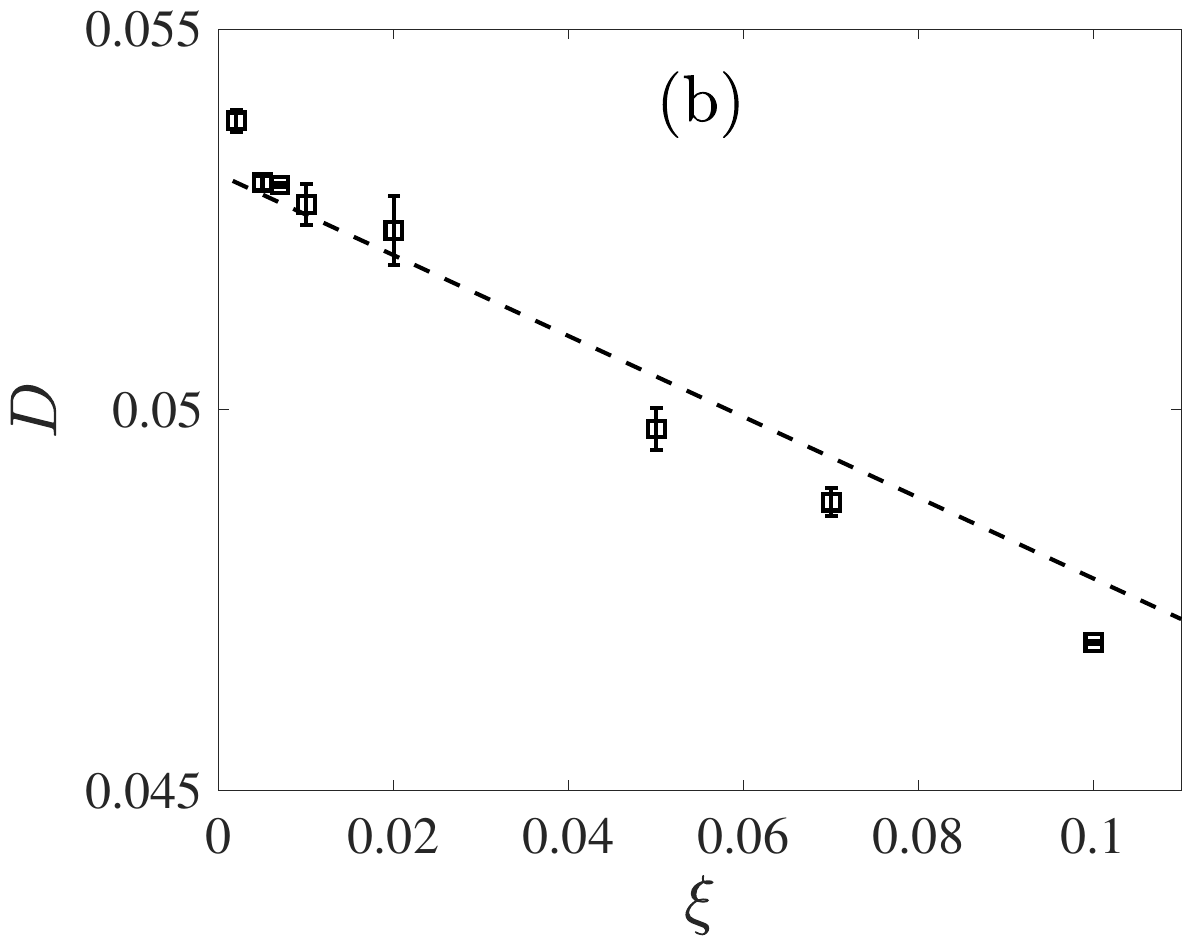}
\caption{(a) Diffusion coefficient $D$ vs $Q$ for regular MPC fluid ($\xi=0$). Top and bottom correspond to temperatures $T=0.5$ and $T=0.1$, respectively. The dashed lines are the theoretical result Eq.\ \eqref{D_Kikuchi}. 
(b) Diffusion coefficient for MPC fluid in porous medium versus friction coefficient $\xi$ for $Q=50$, $\Delta t =1$ and $T=0.1$. The dashed line corresponds to Eq.\ \eqref{D_porous}.}
\label{Diffusion.fig}
\end{center}
\end{figure}

To study self-diffusion under equilibrium conditions, no pressure gradient is applied, $\bF^{\rm ext}=0$. 
Furthermore, to study self-diffusion in an unbounded domain, we consider in this section a periodic system without any walls present. 
Under these conditions and for the two-dimensional angular momentum-conserving collision model adopted here, the diffusion coefficient was calculated using molecular chaos assumption \cite{noguchi_transport_2008} as
\begin{equation} \label{D_Kikuchi}
D = \frac{\kT\Delta t}{m} \left( \frac{1}{\sK} - \frac{1}{2} \right) ,
\end{equation}
with $\sK = 1 - \frac{3}{2Q} + e^{-Q}(3/Q+1-Q/2)/2$. 
Figure \ref{Diffusion.fig}(a) shows the diffusion coefficient as a function of the average number of MPC particles per collision cell $Q$ for two selected temperatures. 

We perform MPC simulations for a two-dimensional periodic system of size $30\times 30$ with 
and different values for $Q$. 
From the particle mean-square displacement, $\ave{[\br_i(t)-\br_i(0)]^2}=4Dt$, we extract the diffusion coefficient $D$ from a least-square fit and verified that the $x$- and $y$-components of the displacement agree with each other within numerical accuracy. 
Figure \ref{Diffusion.fig}(a) 
shows that Eq.\ \eqref{D_Kikuchi} provides a good representation of the numerical results, even though some quantitative discrepancies can clearly be discerned for small up to moderate values of $Q$. For this model, similar deviations from the theoretical predictions have been reported in Ref.\ \cite{noguchi_transport_2008} and attributed to the limitations of the molecular chaos assumption.

Equation \eqref{D_Kikuchi} has been derived for a standard MPC fluid. 
We are not aware of a corresponding result for an MPC fluid in a porous material. 
Heuristically, we can describe the effect of local damping in the MPC scheme for porous materials by an effective time step 
$\delta t$ in the free-streaming step  \cite{matyka_sedimentwater_2017}, 
$\delta t = (1-\xi\Delta t/m)\Delta t$.
We assume that the reasoning presented in Ref.\ \cite{noguchi_transport_2008} leading to Eq.\ \eqref{D_Kikuchi} remains otherwise unaltered. 
In particular, we assume that collisions occur sufficiently fast so that they are not affected by local friction effects. Thus, we suggest an approximation to the MPC diffusion coefficient for spatially homogeneous porous media as 
\begin{equation} \label{D_porous}
D_\xi = \frac{\kT\Delta t(1-\xi\Delta t/m)}{m} \left( \frac{1}{\sK} - \frac{1}{2} \right)
\end{equation}
with the same expression for $\sK$ as given after Eq.\ \eqref{D_Kikuchi}. 
Figure \ref{Diffusion.fig}(b) shows the variation of the effective diffusion coefficient with increasing value of the friction coefficient $\xi$. 
The agreement of the numerical results with Eq.\ \eqref{D_porous} is satisfactory. 
In particular, we find that the effective diffusion coefficient decreases approximately linear with increasing $\xi$ as predicted by Eq.\ \eqref{D_porous}. 

\subsection{Channel flow} \label{channelNewton.sec}

In the following we consider the pressure-driven channel flow of a fluid through a porous medium. 
Within the MPC model, an applied pressure gradient $\dd p/\dd x$ in flow direction is realized by the external force $\bF^{\rm ext}=\fext{\bf e}_x$ acting on every particle, where $\fext=-\rho^{-1}\dd p/\dd x$. 

\begin{figure}[htbp]
\begin{center}
\includegraphics[width=0.45\textwidth]{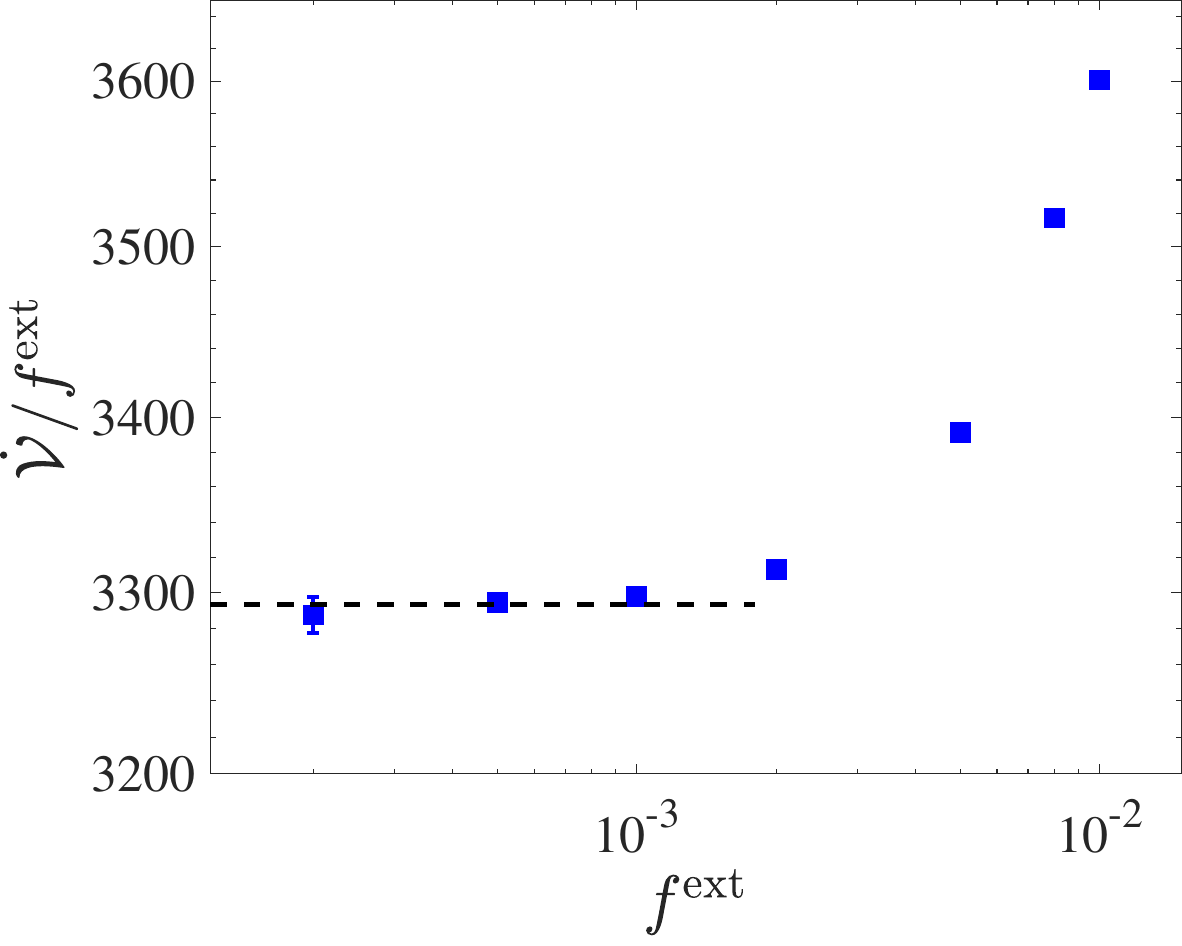} 
\caption{The flow rate divided by the strength of the driving force, $\dot{\mathcal{V}}/\fext$, is shown versus $\fext$ on a semi-logarithmic scale. 
Symbols denote simulation results obtained for parameters $T=0.1$, $\Delta t=0.2$, $Q=100$ and $\xi=0.005$. Dashed line indicates the limiting value for weak driving forces.}
\label{flowrate.fig}
\end{center}
\end{figure}

To determine suitable values for $\fext$, we calculate the flow rate $\dot{\mathcal{V}}$ by numerical integration over the flow profile and verify that this quantity varies linearly with $\fext$ in the parameter regime studied here. 
We consider a planar channel of widths $L=32$ and $64$ and length $50$. 
From Fig.\ \ref{flowrate.fig} we find $\dot{\mathcal{V}}/\fext$ approaches a limiting value with decreasing $\fext$. 
Within statistical uncertainties, we find the same limiting value for $\fext\lesssim 10^{-3}$. 
Therefore, we will use in the following $\fext=10^{-3}$ unless stated otherwise.

Having chosen a suitable value for $\fext$, we perform a series of MPC simulations for different values of the friction coefficient $\xi$ and analyze the resulting velocity profiles. 
For all conditions investigated, we find that the numerical velocity profiles  are well described by the analytical profile for the Darcy-Brinkman model, Eq.\ \eqref{v_channel}. 
Fitting the numerical profiles to  Eq.\ \eqref{v_channel}, we extract two fit parameters, 
$c$ and $\mylambda$, from which the model parameters can be determined as follows. 
First, as shown in Sec.\ \ref{continuum.sec}, $c$ is related to the damping parameter $\alpha$ in the Darcy-Brinkman equation 
\eqref{DarcyBrinkman} by $\alpha=\fext/c$. 
Next, the permeability $K$ is directly linked to the parameter $\mylambda$ by $K=1/\mylambda^2$. 
Finally, the kinematic viscosity is given by 
$\nu=\alpha/\mylambda^2$.  

\begin{figure}[htbp]
\begin{center}
\includegraphics[width=0.45\textwidth]{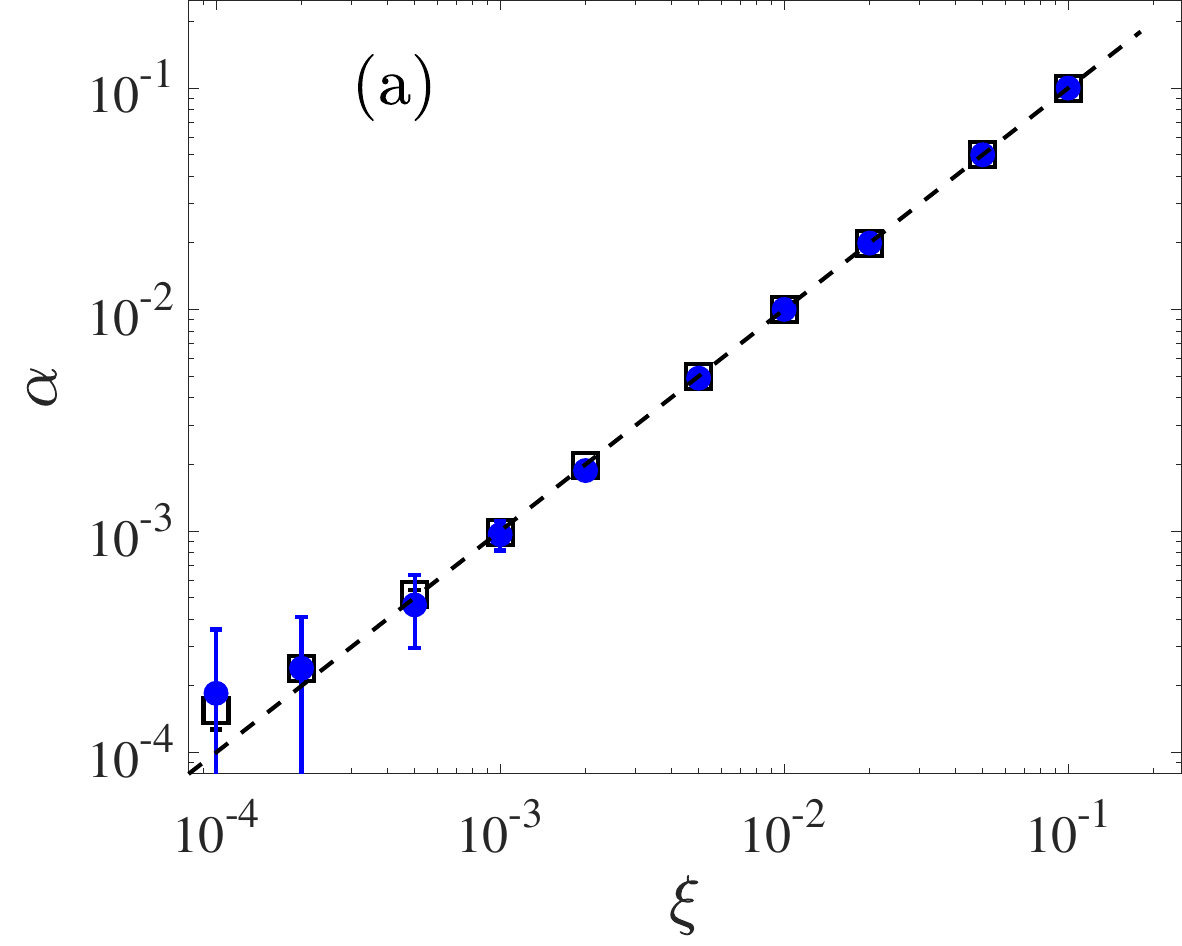}
\includegraphics[width=0.45\textwidth]{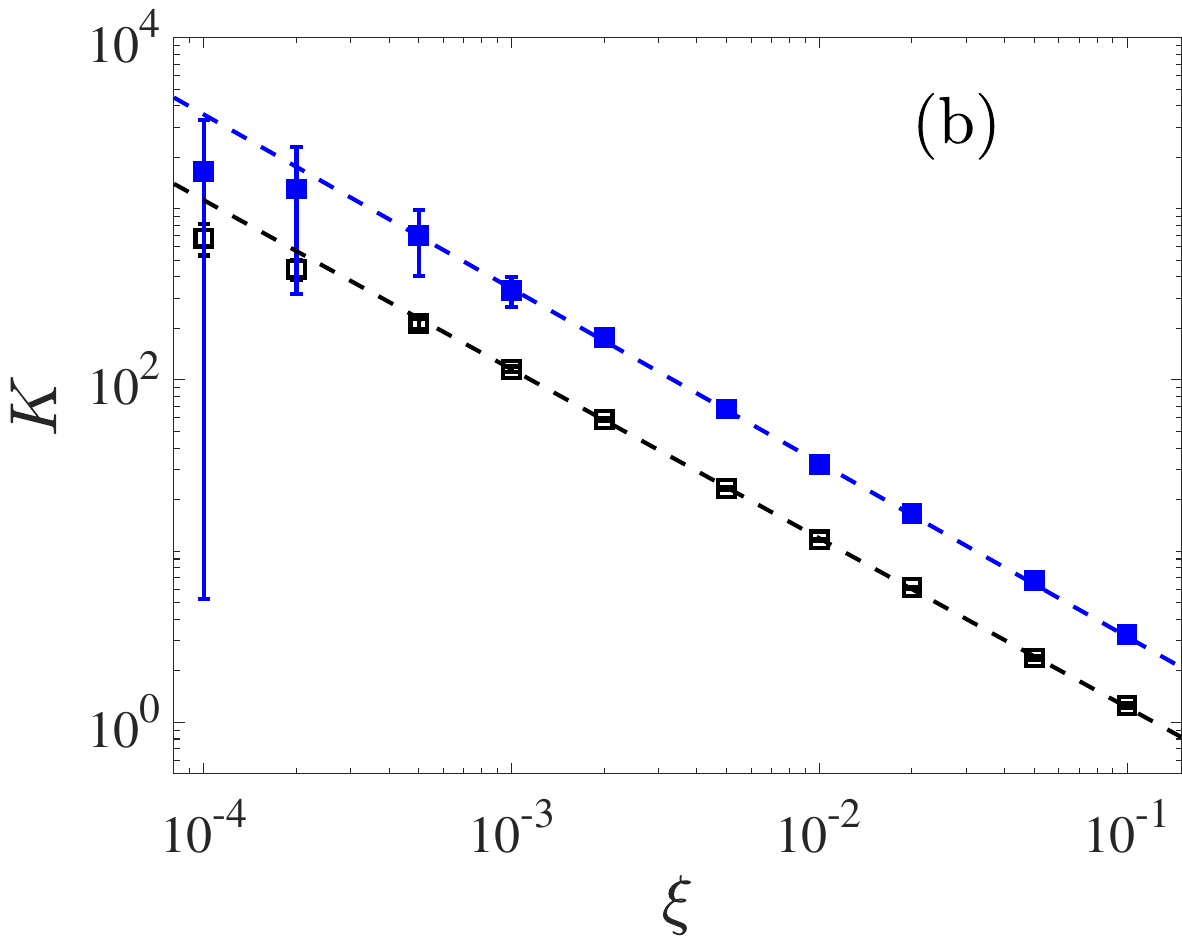}
\caption{(a) The damping parameter $\alpha$ in the Darcy-Brinkman equation \eqref{DarcyBrinkman} extracted from fits to the velocity profile is shown versus the friction coefficient $\xi$ in the MPC model for non-magnetic fluids. Note that a double-logarithmic scale is used. 
Open black squares and filled blue circles correspond to $\Delta t = 1.0$ and $0.2$, respectively. 
The dashed line indicates the relation $\alpha=\xi$. 
(b) Permeability parameter $K$ versus MPC friction coefficient $\xi$ on a double-logarithmic scale. The same model parameters have been chosen and the same symbols are used as in (a). Dashed lines shows the power-law relation $K=k_0\xi^{-\kappa}$ 
with exponent  $\kappa\approx 0.99$ and prefactor $k_0\approx 0.13$ for $\Delta t =1.0$ and 
$\kappa\approx 1.02$ and $k_0\approx 0.30$ for $\Delta t = 0.2$.}
\label{alpha_xi.fig}
\end{center}
\end{figure}

Figure \ref{alpha_xi.fig} shows the extracted damping parameter $\alpha$ and permeability $K$ over a wide range of values for the friction coefficient $\xi$ in the MPC model. 
Within numerical accuracy, we find that the Darcy-Brinkman damping parameter $\alpha$ is equal to the friction coefficient $\xi$ used in the MPC model. 
Therefore, the newly introduced friction coefficient in the MPC model can be identified with the more familiar damping parameter in the Darcy-Brinkman approach. 
For a derivation of this result at least for inviscid fluids see Appendix \ref{DampingDarcy.sec}. 
From  Fig.\ \ref{alpha_xi.fig}(b) we find that the permeability $K$ decreases with increasing $\xi$. 
Except at very small values of $\xi$, we find that the decrease can be described as $K\sim 1/\xi$ to a very good approximation. 
It is interesting to note that the relations $\alpha=\xi$ and $K=k_0\xi^{-1}$ also hold for a corresponding Lattice Boltzmann implementation of flow through porous media \cite{dardis_lattice_1998}, where the density of scatterers plays the role of $\xi$.

For small values of $\xi$, large uncertainties in model parameters extracted from fits to Eq.\ \eqref{v_channel} are found. 
 In this regime, the velocity profiles are close to parabolic, leading to ambiguities in the two-parameter fit to Eq.\ \eqref{v_channel}. 

\begin{figure}[htbp]
\begin{center}
\includegraphics[width=0.45\textwidth]{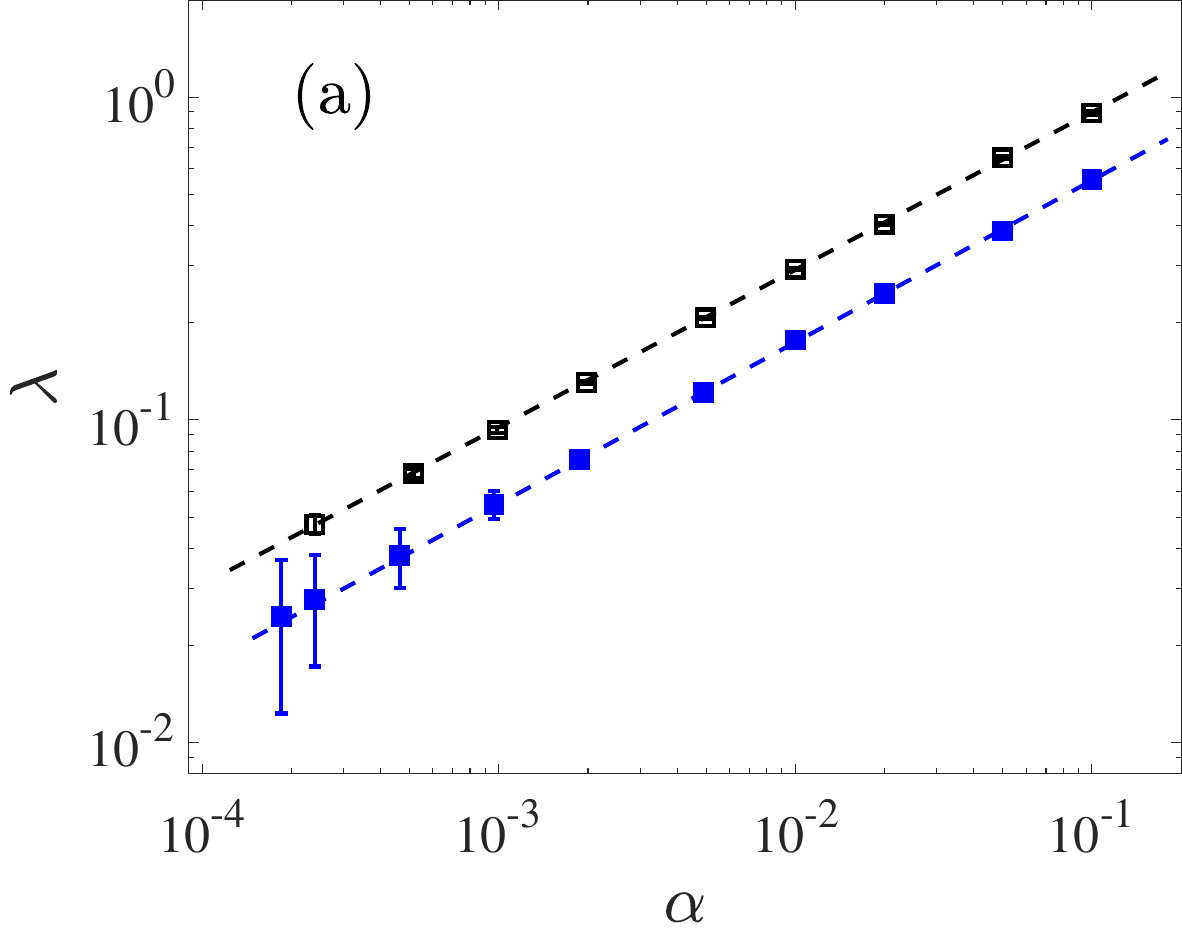}
\includegraphics[width=0.45\textwidth]{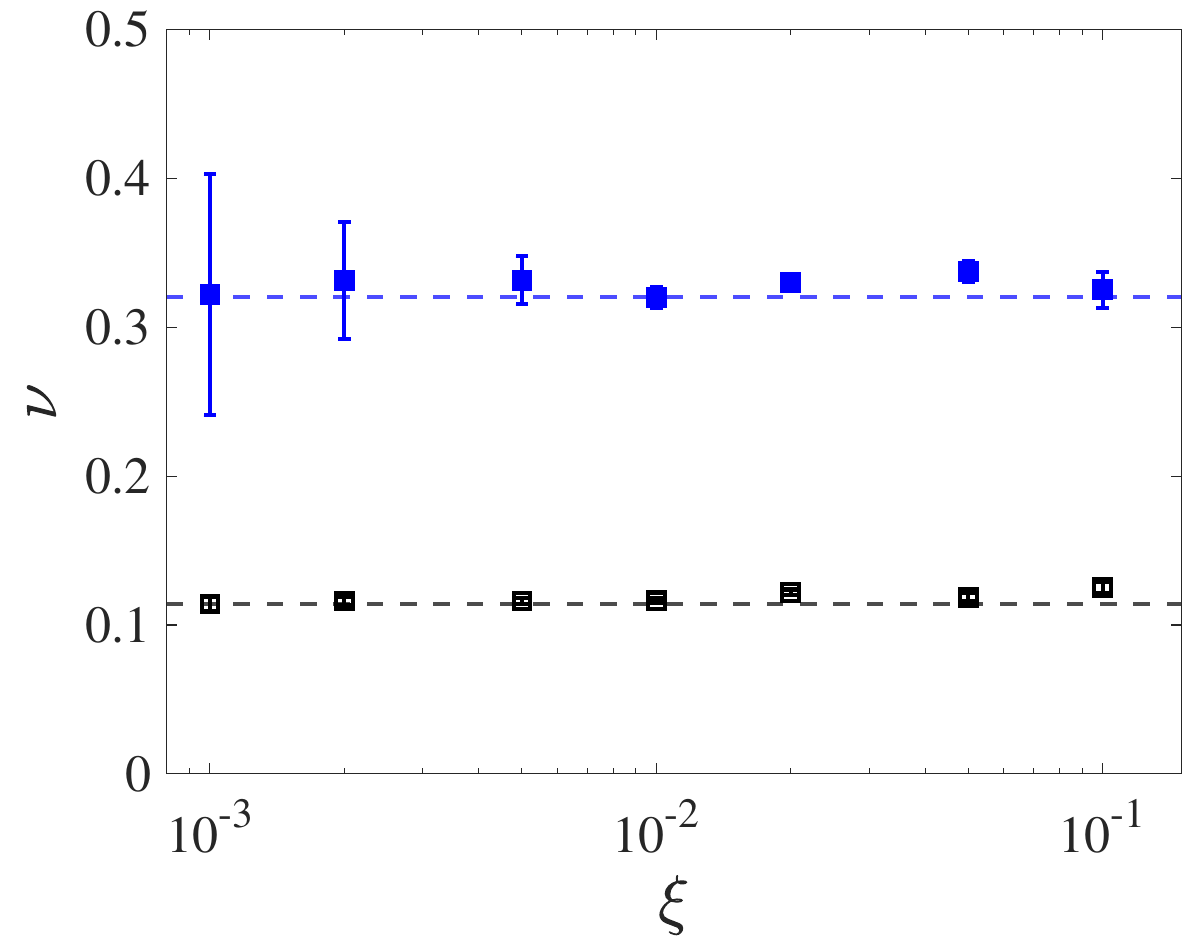}
\caption{(a) The parameter $\mylambda$ is shown parametrically versus $\alpha$ for different friction coefficients and non-magnetic fluids. The same color coding is used as in Fig.\ \ref{alpha_xi.fig}. Dashed lines are power-law fits 
$\mylambda\sim\alpha^b$ with exponents $b\approx 0.49$ and $0.50$ for $\Delta t=1.0$ and $0.2$, respectively. 
(b) The kinematic viscosity $\nu$ extracted from fits to the velocity profile \eqref{v_channel} versus the friction coefficient $\xi$ in the MPC model. 
The same color coding is used as in Fig.\ \ref{alpha_xi.fig}. 
The dashed lines indicate the kinematic viscosity determined in the absence of a porous medium ($\xi=0$).}
\label{nu_xi.fig}
\end{center}
\end{figure}

As a consistency test, we plot in Fig.\ \ref{nu_xi.fig}(a) the parameter $\mylambda$ governing the velocity profile \eqref{v_channel} parametrically versus the damping parameter $\alpha$ that we determined for different values of the friction coefficient. 
In agreement with Eq.\ \eqref{lambdasq} we find from our simulations $\mylambda\sim\alpha^{1/2}$. 
Note that this relation was also confirmed in Lattice Boltzmann simulations \cite{dardis_lattice_1998}, 
while an earlier MPC implementation \cite{matyka_sedimentwater_2017} recovered this relation only over a rather limited interval of $\xi$.

From the relations $\alpha=\xi$ and $K=k_0\xi^{-1}$ extracted from Fig.\ \ref{alpha_xi.fig}, we conclude that the kinematic viscosity $\nu=\alpha K$ (see Sec.\ \ref{continuum.sec}) is given by 
$\nu = k_0$, independent of the friction coefficient. 
In the absence of porous media, we have already established the value of the kinematic viscosity 
$\nu=0.114\pm 0.001$ for $\Delta t=1.0$ in Ref.\ \cite{ilg_multiparticle_2022}. 
Since the current model reduces for $\xi=0$ to a pure MPC fluid, we have performed some 
simulations for $\xi=0$ and fitted the resulting velocity to a parabolic profile expected for Poiseuille flow. 
Thereby, we confirm the value of $\nu$ obtained earlier for the current choice of parameters and $\Delta t =1.0$ and find in addition 
$\nu=0.320\pm 0.001$ for $\Delta t=0.2$.

In Fig.\ \ref{nu_xi.fig}(b) we show the kinematic viscosity $\nu$ obtained from fits to the velocity profile \eqref{v_channel} via the relation $\nu=\alpha/\mylambda^2$ in Sec.\ \ref{continuum.sec}. 
From Fig.\ \ref{nu_xi.fig}(b) we indeed find that $\nu$ is independent of the friction coefficient within statistical uncertainties, consistent with our conclusions above.  
Therefore, we can identify $\nu$ with the kinematic fluid viscosity in the absence of a porous medium.

\section{Ferrofluid fluid flow through porous media}

In this section, we consider ferrofluid flow for the same conditions as studied in Sec.\ \ref{channelNewton.sec}, i.e.\ the steady-state flow through planar channels filled with porous media. 
As before, we assume spatially homogeneous porous materials so that the friction coefficient $\xi$ in the MPC method, Eq.\ \eqref{friction}, is constant, independent of the position throughout the channel. 
Here, we employ the hybrid MPC-BD model of fluctuating ferrohydrodynamics and include the Kelvin-Helmholtz force into the momentum balance as described in Sec.\ \ref{MPCBD}.

Assuming that the rotational relaxation of MNPs is slow compared to fluid motion, we choose $\tauB=100$ and set $\Delta t_{\rm B}=\Delta t$.

\begin{figure}[htbp]
\begin{center}
\includegraphics[width=0.45\textwidth]{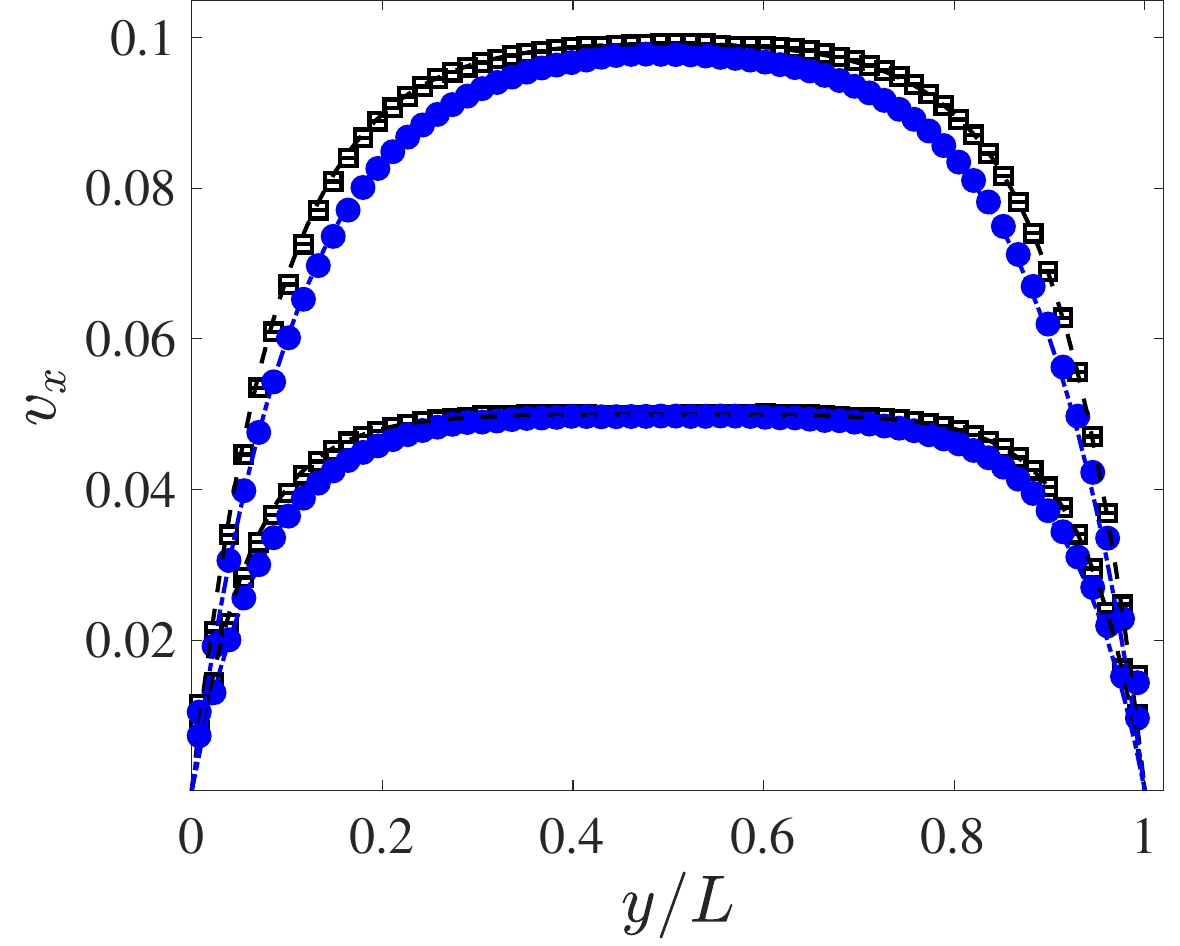}
\caption{The velocity profiles $v_x(y)$ across the channel are shown for ferrofluid flow through a porous medium with $h=0$ (open squares) and $h=5$ (filled circles) for $\Delta t =0.2$, $n^\ast=0.005$. Top curves correspond to $\xi=0.01$, bottom ones to $\xi=0.02$. 
Dashed lines show fits to the profile \eqref{v_channel} in the Darcy-Brinkman model.}
\label{vx_h.fig}
\end{center}
\end{figure}

Figure \ref{vx_h.fig} shows velocity profiles obtained from MPC-BD simulations for selected parameter values of $\Delta t =0.2$, $n^\ast=0.005$, using $\xi=0.01$ and $\xi=0.02$. 
As found earlier, with increasing friction $\xi$, the flow velocity is reduced and deviates more strongly from the  parabolic Poiseuille profile found in the absence of porous media. 
In addition, we observe that the external magnetic field also influences the flow profile. In particular, we find that increasing the magnetic field leads to a reduction of the overall velocity, corresponding to an increase in the effective viscosity of the fluid. This so-called magnetoviscous effect is well-known for magnetic nanoparticles suspended in viscous solvents \cite{Ilg_lnp}. 

Here, we analyze this phenomenon quantitatively for the flow through porous media.  
To this end, we fit the velocity profiles that we obtain numerically from the MPC-BD simulations to 
the profile \eqref{v_channel} resulting from the Darcy-Brinkman model. 
For all parameter values investigated, we find that Eq.\  \eqref{v_channel} provides an accurate description of our numerical results. 
Therefore, we can proceed with our analysis of the fitted coefficients $c$ and $\mylambda$. 
 First, we extract the damping parameter $\alpha$ from the amplitude $c$ of the profile  \eqref{v_channel} via the relation $\alpha=f^{\rm ext}/c$ found in Sec.\ \ref{continuum.sec}, where $f^{\rm ext}=-\rho^{-1}\dd p/\dd x$. 

\begin{figure}[htbp]
\begin{center}
\includegraphics[width=0.45\textwidth]{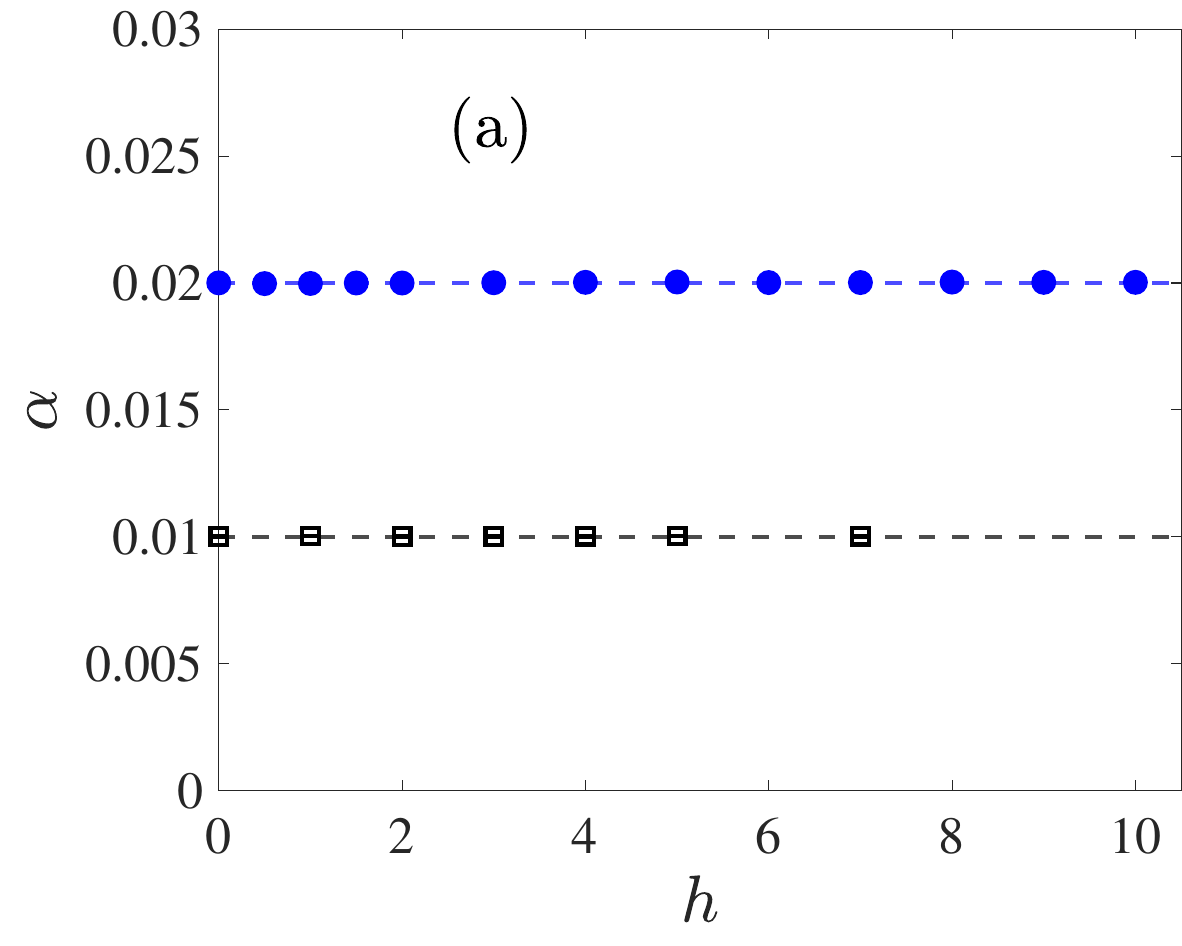}
\includegraphics[width=0.45\textwidth]{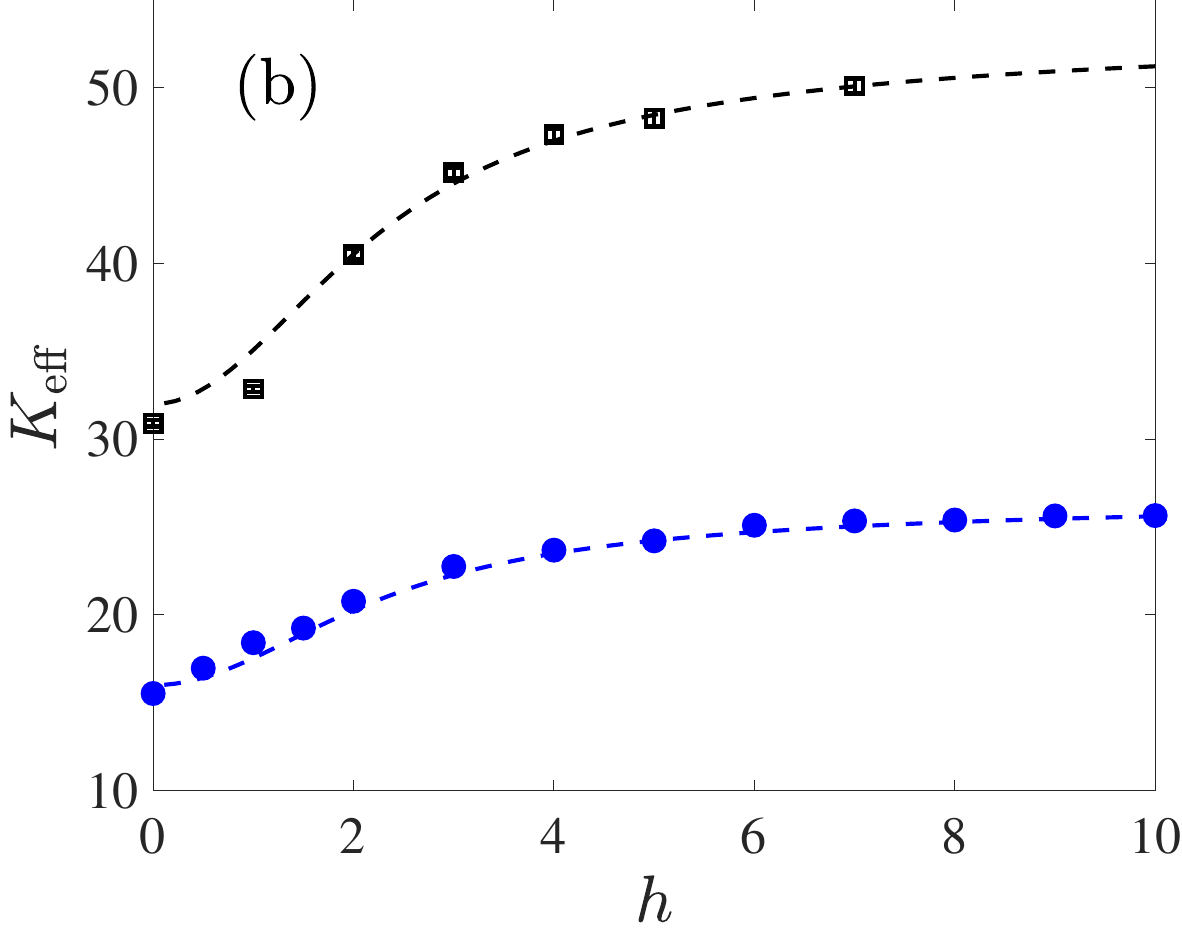}
\caption{(a) The effective damping parameter $\alpha$ in the Darcy-Brinkman equation \eqref{DarcyBrinkman} extracted from fits to the velocity profile is shown versus the applied magnetic field strength $h$. 
The same model parameters are chosen as in Fig.\ \ref{vx_h.fig}
with $\xi=0.01$ (lower) and $\xi=0.02$ (upper). 
(b) Effective permeability parameter $K_{\rm eff}$ versus magnetic field strength $h$. The same model parameters have been chosen and the same symbols are used as in (a). 
Dashed lines show the theoretical result \eqref{Keff_h} explained below.}
\label{alpha_h.fig}
\end{center}
\end{figure}

From Fig.\ \ref{alpha_h.fig}(a) we find that the effective damping parameter $\alpha$ for ferrofluid flow is independent of the magnetic field strength and still given by the MPC friction coefficient $\xi$ as in the non-magnetic/field-free case. 
Thus, within statistical uncertainty, the friction coefficient $\xi$ in the MPC model is equal to the damping parameter $\alpha$ in the Darcy-Brinkman model \eqref{DarcyBrinkman} also in the magnetic case. 
In Fig.\ \ref{alpha_h.fig}(b), we show the effective permeability $K_{\rm eff}$ obtained from the parameter $\mylambda$ in the fitted velocity profile \eqref{v_channel} versus the magnetic field strength $h$ for different values of the friction coefficient $\xi$.  
We observe that $K_{\rm eff}$ increases with increasing field strength, reaching a limiting value for large $h$.

\begin{figure}[htbp]
\begin{center}
\includegraphics[width=0.45\textwidth]{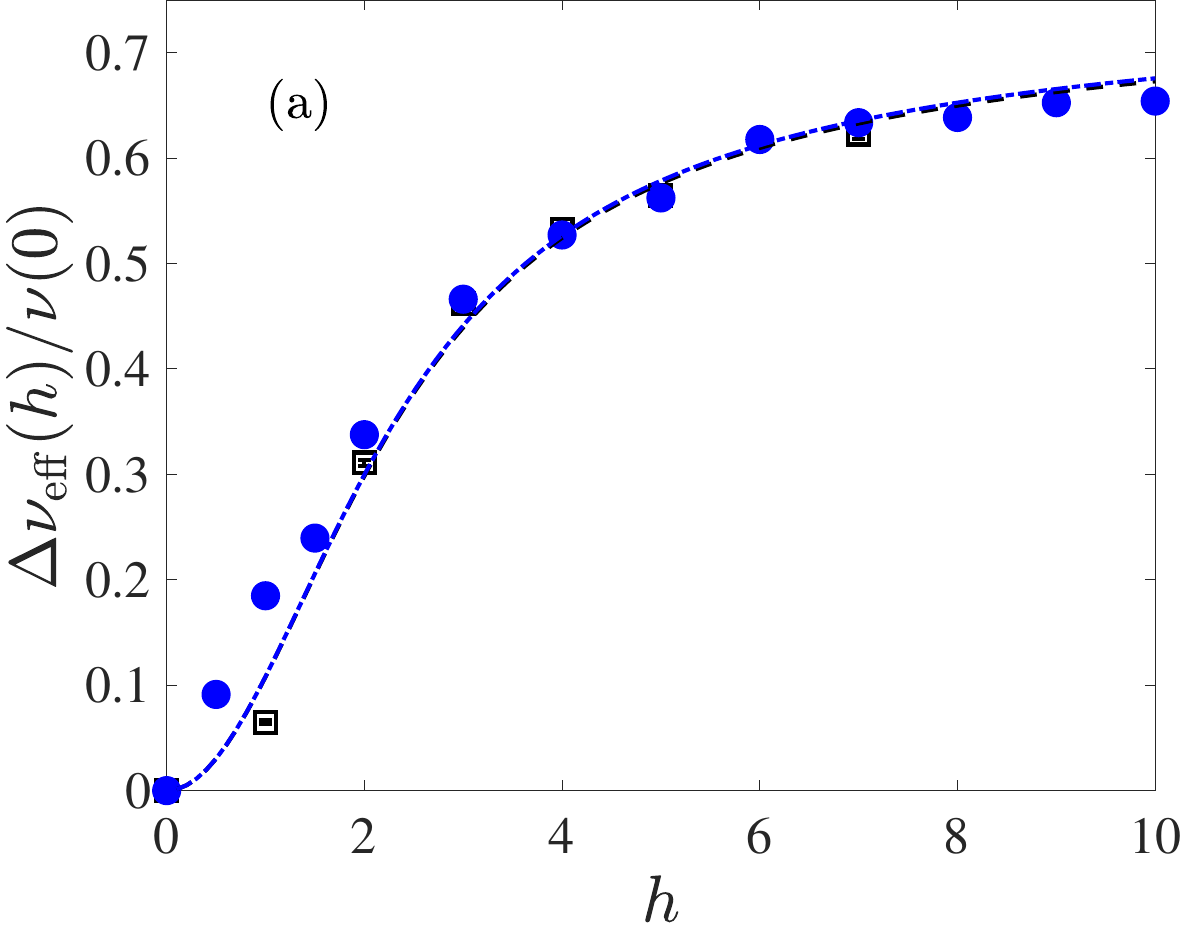}
\includegraphics[width=0.45\textwidth]{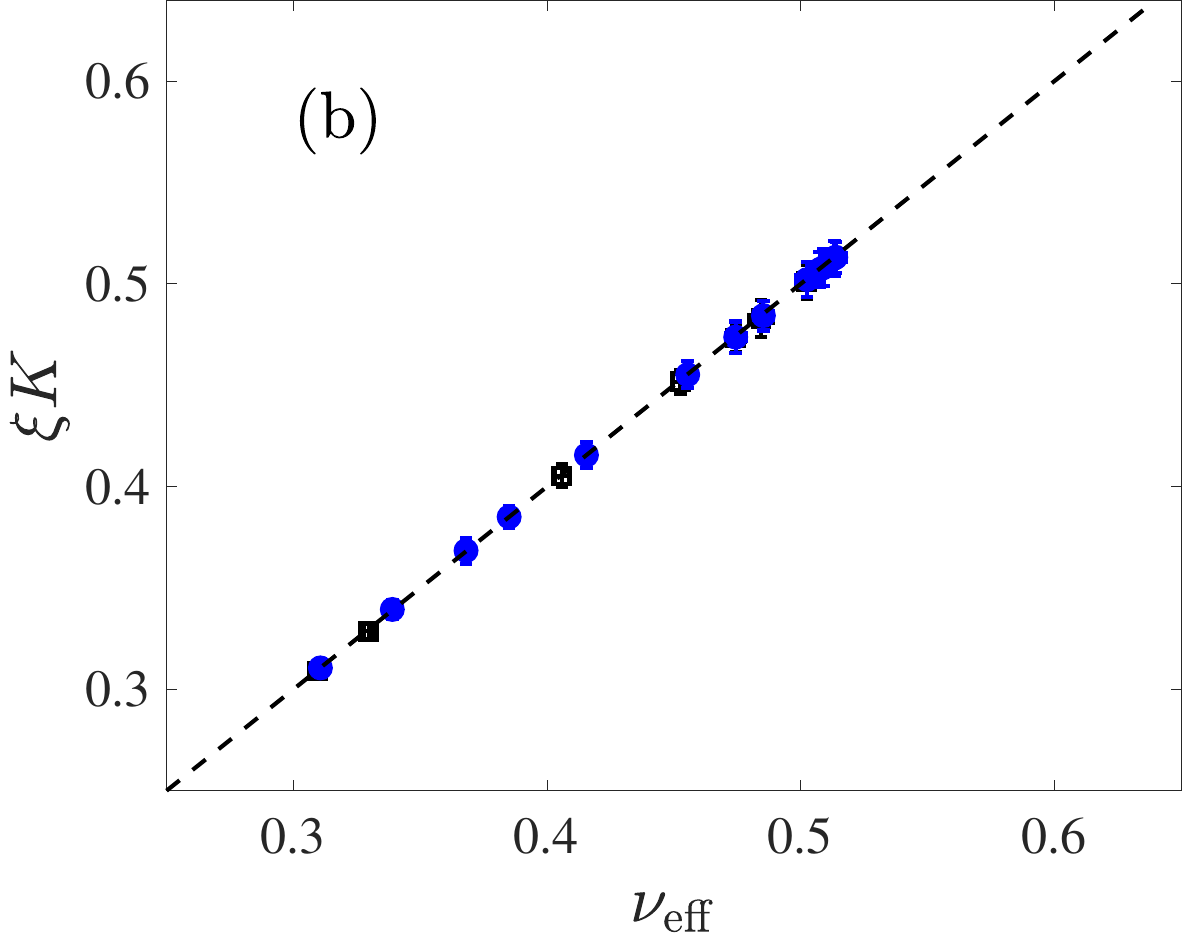}
\caption{(a) The relative change of the effective viscosity, $\Delta \nu_{\rm eff}(h)/\nu(0)$, versus the applied magnetic field strength $h$. 
The model parameters are chosen as in Fig.\ \ref{alpha_h.fig}. 
Dashed line shows a fit to Eq.\ \eqref{nuMRS}. 
(b) The permeability $K$ scaled with the friction coefficient $\xi$ is shown versus the effective viscosity $\nu_{\rm eff}$ for the same data as in panel (a) and in Fig.\ \ref{alpha_h.fig}(b). 
The dashed line shows the relation $\xi K = \nu_{\rm eff}$.}
\label{nueff_h.fig}
\end{center}
\end{figure}

In order to explain this finding, we use  Eq.\ \eqref{lambdasq} to define the effective viscosity $\nu_{\rm eff}$ and evaluate this quantity with the values for the damping parameter $\alpha$ and the fit parameter $\mylambda$ that we have determined. 
In Fig.\ \ref{nueff_h.fig}(a), we show the viscosity change $\Delta\nu_{\rm eff}(h)=\nu_{\rm eff}(h)-\nu(0)$, scaled with the viscosity $\nu(0)$ at zero field. 
We verified that $\nu(0)$ agrees with the viscosity $\nu$ obtained in Sec.\ \ref{channelNewton.sec} for non-magnetic fluids with otherwise identical model parameters. 
From Fig.\ \ref{nueff_h.fig}(a), we find that the relative viscosity change does not depend on the value of the friction coefficient $\xi$. 
Furthermore, the increase of $\Delta\nu_{\rm eff}(h)/\nu(0)$ is well described by the classical result for ferrofluids \cite{MRS74},
\begin{equation} \label{nuMRS}
\frac{\Delta\nu_{\rm eff}(h)}{\nu(0)} = \frac{3}{2}\phi \frac{h L^2(h)}{h-L(h)},
\end{equation}
where $L(h)=\coth(h)-h^{-1}$ denotes the Langevin function and $\phi$ the magnetic volume fraction. 
Therefore, the magnetoviscous effect can be seen in porous media in very much the same manner as in viscous solvent. 

It should be mentioned that the model of Martsenyuk {\em et al.}\   \cite{MRS74}  employed here 
can be justified only for dilute ferrofluids. 
Therefore, in future applications, 
smaller values for the number density $n$ should be chosen with corresponding smaller viscosity changes. 
Here, we have chosen larger values of $n$ to better illustrate the field-dependent effects captured by the simulation method.

Having rationalized the effective viscosity, we replot in Fig.\ \ref{nueff_h.fig}(b) 
the effective permeability $K_{\rm eff}$ from Fig.\ \ref{alpha_h.fig}(b) not versus the magnetic field $h$ but versus $\nu_{\rm eff}(h)$. Multiplying $K_{\rm eff}$ with the friction coefficient $\xi$, all data fall nicely on the diagonal, showing that the effective permeability of ferrofluid flow through porous media is given by 
$K_{\rm eff}(\xi,h) = \nu_{\rm eff}(h)/\xi$, which is well approximated by 
\begin{equation} \label{Keff_h}
K_{\rm eff}(\xi,h)  = \frac{\nu}{\xi}\left[1 +  \frac{3}{2}\phi \frac{h L^2(h)}{h-L(h)}\right],
\end{equation}
with $\nu$ the kinematic viscosity in the absence of magnetic fields. 
In the absence of a magnetic field, we recover $K=K_{\rm eff}(\alpha,h=0)$, where we made use of the equality $\xi=\alpha$. 
For increasing magnetic field strength $h$, we find that the effective permeability $K_{\rm eff}$ increases monotonically to approach an asymptotic value for $h\gg 1$. 
In analogy to the well-known magnetoviscous effect, one might speak of a corresponding ``magneto-permeability effect'' in ferrofluid flow through porous media. 
From Fig.\ \ref{alpha_h.fig}(b) we observe that the magneto-permeability effect for the present model is well-described by Eq.\ \eqref{Keff_h}.

We close this section by considering the most important dimensionless groups determining the nature of fluid flow. 
First, the Reynolds number is defined as ${\rm Re}=UL/\nu$, where $U$ is a characteristic flow velocity and $L$ the channel width. 
For the present choice of parameters $U\approx 0.1$ for $L=64$ (see Fig.\ \ref{vx_h.fig}), so that we find typical Reynolds numbers ${\rm Re}\approx 20$, well in the laminar regime. 
Next, the Schmidt number ${\rm Sc}=\nu/D$ is a measure for the importance of viscous versus molecular diffusion. 
For $Q=100\gg 1$, we find $D\approx T\Delta t/2\approx 0.01$ so that ${\rm Sc}\approx 30$, indicating that we are indeed operating in the relevant regime for fluids, where collisions dominate over kinetic transport. 
Finally, the Mach number is defined as ${\rm Ma}=U/c_{\rm s}$, where $c_{\rm s}=\sqrt{5\kT/3m}$ denotes the speed of sound. 
In our case, we find typically ${\rm Ma}\approx 0.25$. 
It is well-known that particle-based methods like MPC do not strictly obey the incompressibility condition. 
The smaller the Mach number, the better incompressibility is restored. 
Finally, for viscoelastic fluids like ferrofluids, the Weissenberg number ${\rm Wi}=\tauB U/L$ gives the ratio of viscous to elastic forces. Here, we typically find ${\rm Wi}\approx 0.2$ and therefore the simulations are performed in the Newtonian regime.

\section{Flow through channels with walls covered by porous media}

In this section, we consider again driven flow through a parallel channel. 
This time, however, the porous medium is only present within a layer of width $\ellp$ on both walls, 
whereas the fluid is unperturbed in center. 

Within the MPC scheme described in Sec.\ \ref{MPCbasics.sec}, this situation can be implemented conveniently by 
a position-dependent friction coefficient $\xi(\br)$ in Eq.\ \eqref{friction}. 
The porous layers are described by setting 
$\xi(\br)=\xi$ for $y\leq\ellp$ or $y\geq L-\ellp$ and $\xi=0$ in the center, $\ellp < y < L-\ellp$.

\begin{figure}[htbp]
\begin{center}
\includegraphics[width=0.45\textwidth]{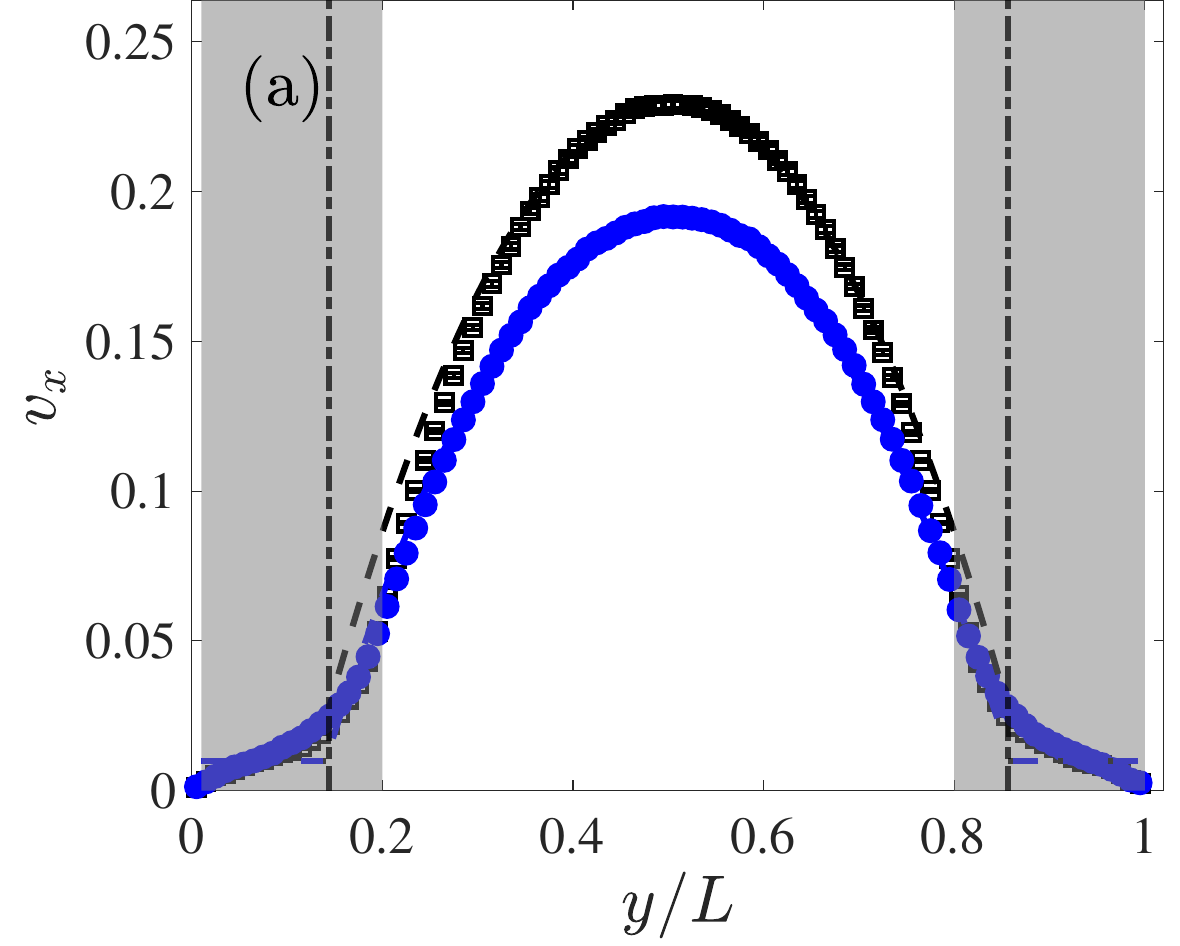}
\includegraphics[width=0.45\textwidth]{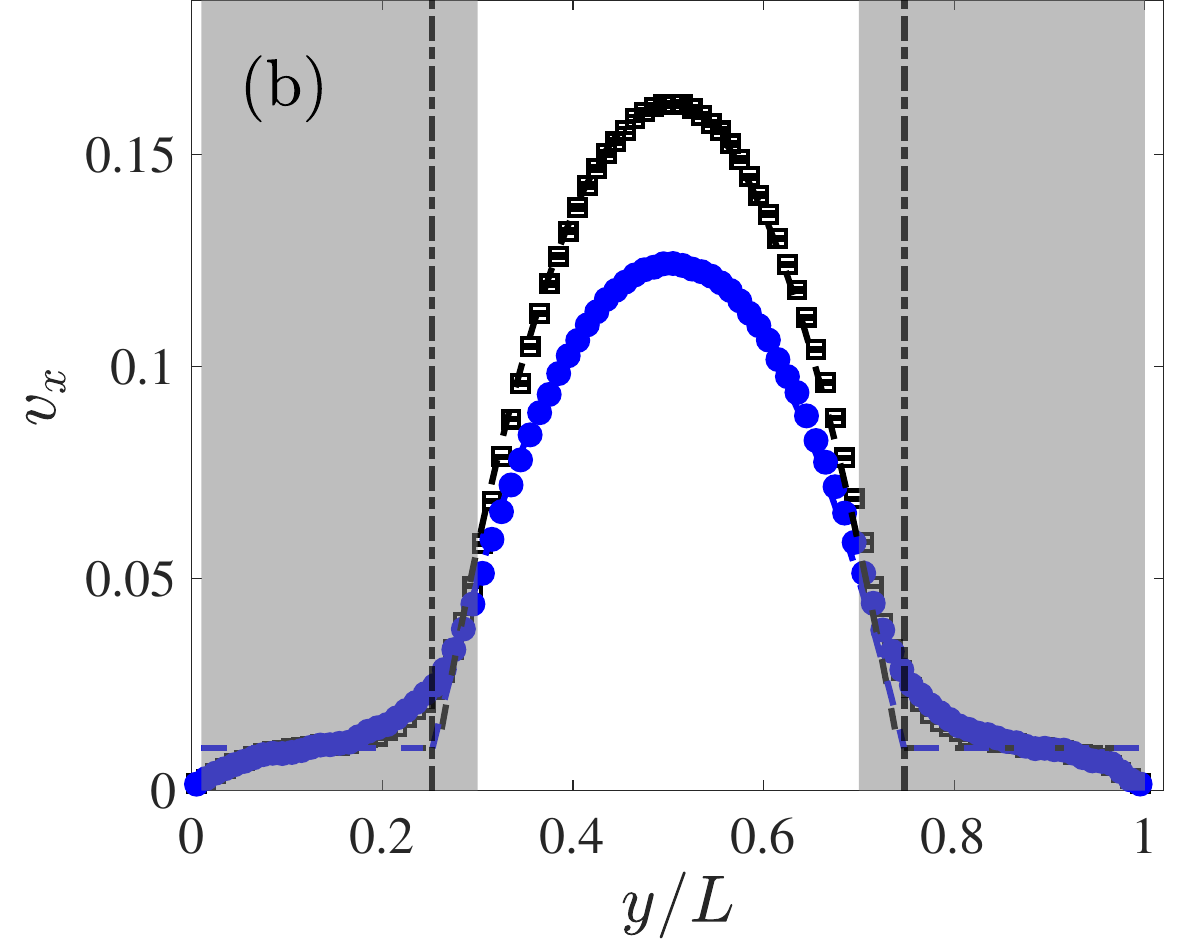}
\caption{Flow profile in a parallel channel where walls are covered with a porous layer, indicated by the grey shaded regions. 
Black and blue symbols correspond to $h=0$ and $h=5$, respectively.  
In panels (a) and (b), the width of the porous layer is $\ellp = 0.2 L$ and $0.3 L$, respectively. 
The model parameters are chosen as $T^\ast=0.1$, $\Delta t=0.2$, $Q=100$, $\xi=0.02$, $\fext=0.0002$. 
The dot-dashed lines indicate the estimated value for $\ellp$ based on Eq.\ \eqref{dotVfromellp} and dashed lines show the corresponding simplified profile \eqref{v_layers}.}
\label{v_profiles_layers.fig}
\end{center}
\end{figure}

Figure \ref{v_profiles_layers.fig} shows typical simulation results of the velocity profile obtained in such a situation. 
From Fig.\ \ref{v_profiles_layers.fig} we observe a smooth transition between a parabolic Poiseuille profile in the center and a more flat Darcy-Brinkman profile in the porous layers. 
These observations are in qualitative agreement with earlier simulations \cite{matyka_sedimentwater_2017}. 
In the center region, the magnetic field increases the effective viscosity, therefore slowing down the flow. 
In the porous region, however, the magnetic field is found to have a much smaller effect on the flow. 

In a simplified description, we can approximate the velocity as constant, $v=v_0$, within the porous layers  of width $\ellp$, and a parabolic profile in the center of the channel. 
Insisting on a continuous velocity profile at the interface between the two at $y=\ellp$ and $y=L-\ellp$, we arrive at  
\begin{equation} \label{v_layers}
v_x(y) \approx \left\{ 
\begin{array}{cl}
v_0 & {\rm for}\ y<\ellp\ {\rm or}\ y>L-\ellp\\
v_0 + u[y(L-y) - \ellp(L-\ellp) ]&{\rm for}\  \ellp\leq y \leq L-\ellp
\end{array}\right.
\end{equation}
with $L$ the channel width. 
The quantity $u$ can be expressed in terms of the maximum velocity $v_{\rm max}$ in the channel center as 
$u=(v_{\rm max}-v_0)/[(L/2)^2-\ellp(L-\ellp)]$. 
The flow rate $\dot{\mathcal{V}}=\int_0^Lv_x(y)\dd y$ 
for the flow profile \eqref{v_layers} can be calculated as 
\begin{equation} \label{dotVfromellp}
\dot{\mathcal{V}} 
=  \frac{1}{3}(2v_{\rm max}+v_0)L - \frac{4}{3}(v_{\rm max}-v_0)\ellp.
\end{equation}
For constant $v_0<v_{\rm max}$, increasing the width $\ellp$ of the porous layer leads to a proportional decrease in the flow rate.

In view of possible applications, it might be useful to be able to estimate the thickness of the porous layer $\ellp$ without the need to determine the detailed flow profile. 
We assume one can measure the total flow rate $\dot{\mathcal{V}}$ and the centerline velocity $v_{\rm max}$. 
In our simulations, we obtain $v_{\rm max}$ from fitting a parabola to the flow profile in the center region. 
Furthermore, we approximate the velocity within the porous layer by Darcy's law  \eqref{Darcy}, 
$v_0 = -(K/\eta)\dd p/\dd x = \fext/\xi$.
Knowing the channel width $L$, we can then determine the thickness of the porous layer within the simple model \eqref{v_layers} by solving Eq.\  \eqref{dotVfromellp} for $\ellp$. 

Vertical dot-dashed lines in Fig.\ \ref{v_profiles_layers.fig} indicate the value of $\ellp$ obtained in this way, while dashed lines show the corresponding approximate profile \eqref{v_layers}. 
Note that changing the magnetic field strength alters the flow profile, but the estimate for $\ellp$ remains unchanged within numerical accuracy. 
From Fig.\ \ref{v_profiles_layers.fig}, it is apparent that Darcy's law captures the mean velocity, but approximating the flow as constant within the porous layer is a rather crude approximation. Consequently, the layer thickness is underestimated by around 15-30\% within the parameter range investigated. In spite of this inaccuracy, such a simple estimate might be a useful first step in determining the thickness of porous layers.

\section{Conclusions}

In this communication, we propose an extension of the hybrid MPC-BD scheme developed in Ref.\ \cite{ilg_multiparticle_2022} to describe fluctuating ferrohydrodynamics in porous media by additional friction forces. 
We performed numerous computer simulations and validated the model and its implementation in several ways and over a considerable range of parameters. 
In particular, we verified that the newly introduced friction coefficient is identical to the damping parameter in the Darcy-Brinkman model. 
Using an Irving-Kirkwood approach, we argue that this identity is expected, at least on large enough scales where hydrodynamics emerges from the MPC model. 
We also verified the expected dependence of the permeability on the friction coefficient. 
Therefore, the current MPC model can serve as an alternative to the Lattice-Boltzmann implementation proposed in Ref.\ \cite{dardis_lattice_1998} for the non-magnetic case.
We mention that our implementation of the MPC method for non-magnetic fluids improves on an earlier approach \cite{matyka_sedimentwater_2017}, which suffered from a number of misconceptions concerning model parameters. 

In the present work, we benefit from the flexibility of the MPC approach and couple the stochastic rotational motion of MNPs to the MPC scheme for flow through porous media. 
From simulations of channel flows, we find a ``magneto-permeability effect'' in porous media, i.e.\ the field-dependent change of the effective permeability.  
This effect is analogue to the traditional magnetoviscous effect in  ferrofluids \cite{Ilg_lnp} and can be described theoretically in the same manner.  

As an application of the method, we consider the flow through a planar channel with walls  covered by a layer of porous material. 
In this situation, a parabolic velocity profile develops in the center of the channel, as is well-known for Poiseuille flow of viscous fluids. On approaching the porous layers, the parabolic profile gives way to the more plug-like flow, which is typically observed in porous materials. 
We propose a simple method to estimate the thickness of the porous layer, based only on the total flow rate and the centerline velocity. 
Such rough estimates might be useful in a number of practical applications.

The present study can be extended in various ways. 
First, the extension to fully three-dimensional flows and more complicated geometries is straightforward, thanks to the flexibility of the MPC method. 
These extensions, together with a generalized model to describe non-dilute ferrofluids 
have already been explored for flow in non-porous media \cite{ilg_simulating_2022}. 
From a more conceptual point of view, future studies are needed to investigate the role of porous media on the rotational motion of nanoparticles. Once these effects are better understood, they can then be included within the present hybrid MPC-BD scheme. 
A further extension of the present work could be the investigation of strongly heterogeneous, e.g.\ fractured porous media and to revisit earlier studies \cite{huang_numerical_2021} with a more reliable constitutive model. 
The current approach is also well-suited to be incorporated within multi-scale simulation schemes  \cite{boccardo_review_2020,al_sariri_multi-scale_2022}, which are needed e.g.\ 
to study the highly interesting phenomenon of colloidal deposition in porous media \cite{bizmark_multiscale_2020,gerber_propagation_2020}.

\begin{appendices}

\section{Friction and Damping} \label{DampingDarcy.sec}

The MPC equations for the streaming step \eqref{r_streaming}, \eqref{v_streaming} can be interpreted as a discretization of their continuum counterpart 
\begin{align} \label{dt_r}
\dot{\br}_i & = \bp_i/m_i \\
\dot{\bp}_i & = -\xi\bv_i + \bF^{\rm ext}. \label{dt_p}
\end{align}
We restrict the following discussion to the inviscid limit and neglect collisions, which corresponds to neglecting inter-particle interactions. 

Following the classical Irving-Kirkwood procedure \cite{IrvingKirkwood}, we define hydrodynamic fields by suitable averages in terms of particle configurations. 
In particular, the mass density can be defined by 
$\rho(\br;t)=\ave{\sum_{i=1}^Nm_i\delta(\br-\br_i(t))}$ and the momentum density by 
$\bg(\br;t)=\ave{\sum_{i=1}^N\bp_i\delta(\br-\br_i(t))}$, with the Dirac delta function $\delta(\br)$. 
Through these definitions, the partial derivative $\partial\bg/\partial t$ can be expressed as 
\begin{equation}
\frac{\partial\bg}{\partial t} = \ave{\sum_{i=1}^N\dot{\bp}_i\delta(\br-\br_i(t))} + \ave{\sum_{i=1}^N\bp_i\dot{\br}_i\cdot\frac{\partial}{\partial\br_i}\delta(\br-\br_i(t))}.
\end{equation}
Inserting Eqs.\ \eqref{dt_r}, \eqref{dt_p} 
and relating $\bF^{\rm ext}$ to the pressure gradient $\bnabla p$ we find 
\begin{equation} \label{IK}
\frac{\partial}{\partial t}\bg = - \frac{\xi}{m}\bg  - \bnabla p - \bnabla\cdot\bsigma^{\rm kin},
\end{equation}
with $\bsigma^{\rm kin} =\ave{\sum_{i=1}^Nm^{-1}\bp_i\bp_i\delta(\br-\br_i(t))}$. 
For simplicity we assumed $m_i=m$. 
For a Newtonian fluid, the kinetic stress tensor is given by $\bsigma^{\rm kin} = -(\eta^{\rm kin}/2)[\bnabla\bv + (\bnabla\bv)^T]$. 
Including collisions will add collisional contributions to the stress tensor but will otherwise leave Eq.\ \eqref{IK} unchanged. 
Relating the momentum density to the velocity field, $\bg=\rho\bv$, and considering the stationary state, we find that Eq.\ \eqref{IK} agrees with Eq.\ \eqref{DarcyBrinkman}, where the damping parameter $\alpha$ is given by $\xi/m$. 
Therefore, in the hydrodynamic limit, the MPC model equations including friction forces recover the Darcy-Brinkman model.

\end{appendices}


\end{document}